\documentclass[aps,prd,floatfix,10pt,nofootinbib,showpacs,notitlepage,twocolumn,superscriptaddress]{revtex4-1}
\usepackage[utf8]{inputenc}

\usepackage{amsmath}
\usepackage{amsfonts}
\usepackage[colorlinks=true,linkcolor=blue,citecolor=blue,urlcolor=blue]{hyperref}
\usepackage{amssymb}
\usepackage{graphicx}
\usepackage{enumerate}
\usepackage{csquotes}
\usepackage{tabularx}
\usepackage{bm}
\usepackage{commath}

\newcommand{\nn}{\nonumber \\}
\newcommand{\eq}[1]{Eq.~(\ref{#1})}
\newcommand{\fig}[1]{Fig.~\ref{#1}}
\newcommand{\Fig}[1]{Fig.~\ref{#1}}

\newcommand{\bsub}{\begin{subequations}}
\newcommand{\esub}{\end{subequations}}
\newcommand{\be}{\begin{eqnarray}}
\newcommand{\ee}{\end{eqnarray}}
\newcommand{\om}{\ensuremath{\omega}}
\usepackage{float}

\newcommand{\pad}{\ensuremath{\partial}}

\newcommand{\lp}{\ensuremath{\left(}}
\newcommand{\rp}{\ensuremath{\right)}}
\newcommand{\bi} {\begin{itemize}}
\newcommand{\ei} {\end{itemize}}
\newcommand{\ben}{\begin{enumerate}}
\newcommand{\een}{\end{enumerate}}
\newcommand{\bmat}{\begin{pmatrix}}
\newcommand{\emat}{\end{pmatrix}}
\newcommand{\eqa}[1]{\begin{align}#1\end{align}}

\begin{document}

\preprint{???}
\title{Slow Sound in a duct, effective transonic flows and analogue black holes \\}

\author{Yves Aur\'egan} \email{yves.auregan@univ-lemans.fr}
\affiliation{Laboratoire d'Acoustique de l'Universit\'e du Maine,\\UMR CNRS 6613, Av. 
O Messiaen, 72085 Le Mans Cedex 9, France}
\author{Pierre Fromholz} \email{pierre.fromholz@ens.fr} \affiliation{D\'epartement de Physique de l'ENS, 24 rue Lhomond, 75005 Paris, France}
\author{Florent Michel}  \email{florent.michel@th.u-psud.fr}
\affiliation{Laboratoire de Physique Th\'eorique, CNRS UMR 8627,\\ B\^at. 210, Universit\'e Paris-Sud 11, 91405 Orsay Cedex, France}
\author{Vincent Pagneux} \email{vincent.pagneux@univ-lemans.fr}
\affiliation{Laboratoire d'Acoustique de l'Universit\'e du Maine,\\UMR CNRS 6613, Av. 
O Messiaen, 72085 Le Mans Cedex 9, France}
\author{Renaud Parentani} \email{renaud.parentani@th.u-psud.fr} 
\affiliation{Laboratoire de Physique Th\'eorique, CNRS UMR 8627,\\ B\^at. 210, Universit\'e Paris-Sud 11, 91405 Orsay Cedex, France}
\date{\today}

\begin{abstract}
We propose a new system suitable for studying analogue gravity effects, consisting of a gas flowing in a duct with a compliant wall. Effective transonic flows are obtained from uniform, low Mach number flows through the reduction of the one-dimensional speed of sound induced by the wall compliance. We show that the modified equation for linear perturbations can be written in a Hamiltonian form. We perform a one-dimensional reduction consistent with the canonical formulation, and deduce the analogue metric along with the first dispersive term. In a weak dispersive regime, the spectrum emitted from a sonic horizon is numerically shown to be Planckian, and with a temperature fixed by the analogue surface gravity. 
\end{abstract}

\maketitle

Engineering flows that are transonic and regular offers the possibility to test well known predictions concerning astrophysical black holes~\cite{Unruh:1980cg}. Of particular interest is Hawking's discovery that black 
holes should spontaneously emit a steady thermal flux~\cite{Hawking:1974sw}. Although this effect was originally phrased in the context of quantum relativistic fields, it rests on the {\it anomalous} mode mixing occurring near the black hole horizon~\cite{Brout:1995rd}. This mixing, which is stationary and conserves the wave energy, is said anomalous as it leads to a mode amplification and involves negative energy waves. Because of the precise analogy between the equation governing sound propagation and that used by Hawking, these key elements are recovered in a stationary transonic flow. Indeed, in the acoustic approximation, for long wavelengths, the mode mixing possesses the main properties of the one responsible for the Hawking effect~\cite{Unruh:1994je,Brout:1995wp}. 

To complete the comparison, one should take into account the dispersive properties of sound waves, which have no counterpart in general relativity. (Note however that dispersive terms appear in certain theories of modified gravity where Lorentz invariance is broken~\cite{Horava:2009uw,Jacobson:2010mx,Coutant:2014cwa}.) Analytical and numerical studies have established that the correspondence is quantitatively preserved provided the two relevant scales are well separated~\cite{Unruh:1994je,Brout:1995wp,Macher:2009tw,Coutant:2011in,Finazzi:2012iu,Robertson:2012ku}, namely when the dispersive length is sufficiently smaller than the typical length scale associated with the inhomogeneity of the flow (which then plays the role of the inverse surface gravity of the black hole). Therefore, there is no conceptual obstacle preventing to test the Hawking prediction by observing the mode mixing across a sonic horizon. In practice, the difficulty is to find appropriate set-ups. Many have been proposed, involving for instance ultra cold atomic clouds~\cite{Garay:1999sk}, surface waves in flumes~\cite{Schutzhold:2002rf}, and light in non-linear media~\cite{Philbin07032008}. Recently, the first experiments have been carried out~\cite{Rousseaux:2007is,Weinfurtner:2010nu,2010PhRvL.105x0401L,BHLaser-Jeff}. 

\begin{figure}
\includegraphics[width=\linewidth]{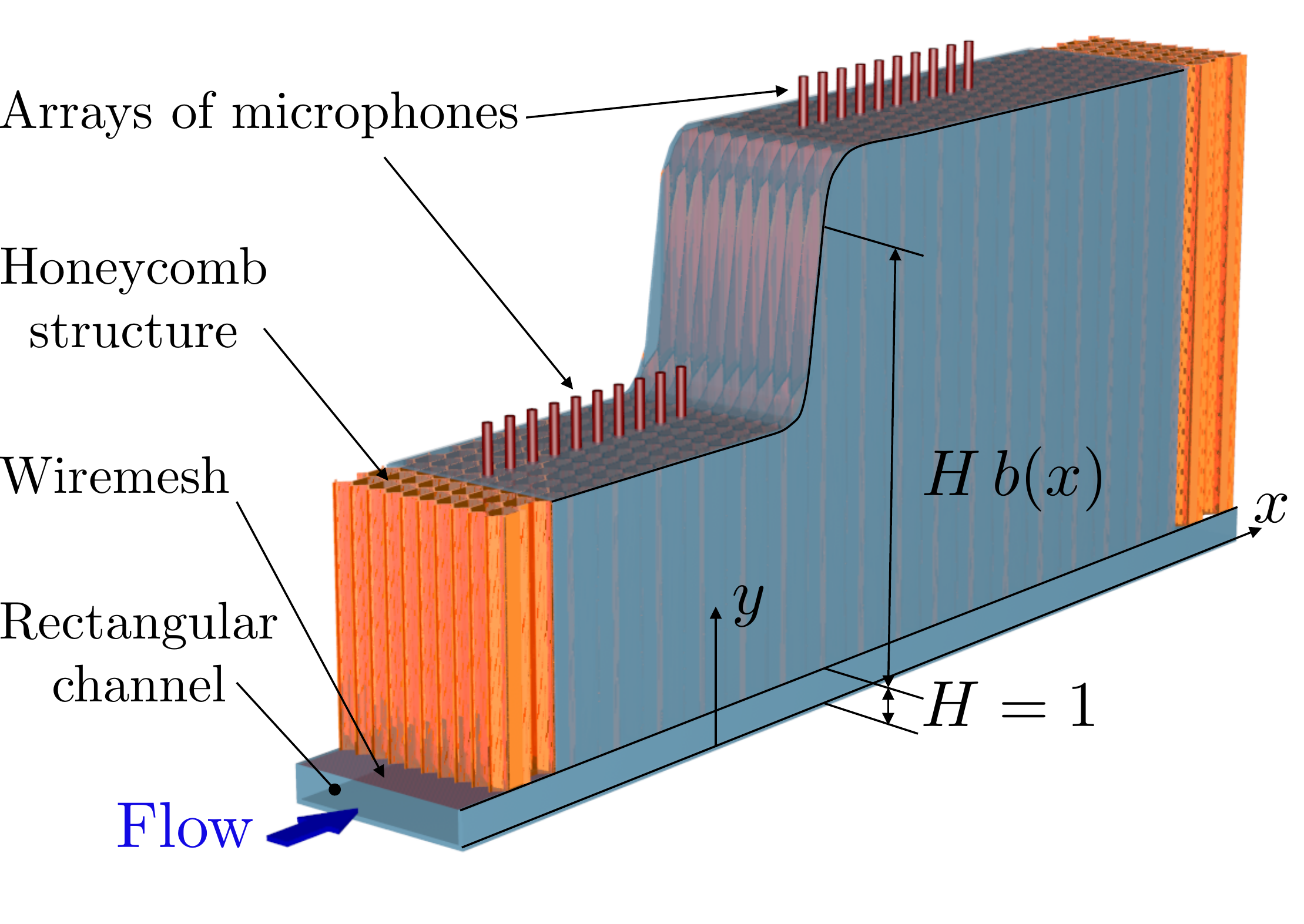}
\caption{Schematic drawing of the configuration. At $y=0$, the wall is rigid. At $y=1$, the compliant wall is made of a succession of tiny tubes of height $b(x)$. As explained in the text, the effective sound speed is a decreasing function of $b$. In the present profile $b(x)$ and for a given uniform flow 
with Mach number $M<1$, an effective supersonic region can be created in the right region.} \label{fig:bx}
\end{figure} 

In this Letter, we propose a new framework which is a variant of the original one~\cite{Unruh:1980cg,Unruh:1994je}. Its main interest resides in obtaining a large reduction of the low frequency one-dimensional sound speed in a duct so that a stationary flow with a uniform low Mach number $M$ possesses a horizon. In realistic settings, $M$ could be close to $0.3$. The reduction of the effective sound speed is achieved by means of a compliant wall composed of thin tubes which modify the upper boundary condition, see Fig.~\ref{fig:bx}. Interestingly, for near-critical flows, the wave equation follows from a well-defined action principle. From this, we derive the acoustic black hole 
metric in the long wavelength regime, and a conserved norm which establishes the anomalous character of the mode mixing. At the end of this Letter we briefly present the practical advantages of this proposal.

{\it The model} --
We consider the propagation of sound waves in a 2-dimensional horizontal channel of uniform height $H$. $x^*$ denotes the cartesian horizontal coordinate, $y^*$ the vertical one, and $t^*$ the time. We assume the flow of air is uniform, with a horizontal velocity $U_0$. Denoting by $c_0$ the sound speed, $\rho_0$ the air density, $\mathbf{v}^*$ the velocity perturbation, and $p^*$ the pressure perturbation, the time evolution is given by \eqa{c_0^{-1} D_t^*p^* = -\rho_0 \nabla^* \cdot \mathbf{v}^*, 
\, 
\; \rho_0 D_t^*\mathbf{v}^* = - \nabla^* p^*, \label{eq:f}
}
where $D_t^* \equiv \pad_{t^*} + U_0 \pad_{x^*}$ is the convective derivative. We define dimensionless quantities as $x = x^*/H$, $y = y^*/H$, $t = t^* c_0 / H$, $\mathbf{v} = \mathbf{v}^*/c_0$, $p = p^*/(\rho_0 c_0^2)$, and $M=U_0/c_0$. The potential $\phi$ gives the velocity by $\mathbf{v} = \nabla \phi$, and the pressure by $p = -D_t \phi$. It obeys
\eqa{\label{eqmouv}
D_t^2\phi- (\partial_x^2 + \partial_y^2)
\phi &= 0.}

At the lower wall, the impenetrability condition is simply $\partial_y \phi(t,x,y=0) = 0$, see Fig.~\ref{fig:bx}. At the upper wall $y=1$, the continuity of the displacement and pressure gives rise to a non local expression in time, 
see~\cite{1980JSV....71..429M,JASA, suppl} for details. However, for near-critical flows and small frequencies, it can be written as 
\eqa{\partial_y \phi + D_t ( b(x) D_t \phi) =0 \; 
\mathrm{at} & \; y=1,  \label{CL2}}
which is second-order in $\pad_t$. 

For a homogeneous stationary flow, we can look for solutions of the form $\varphi_k \propto \cosh(\alpha_k y) e^{i \lp k x - \omega_k t \rp}$. Eq.~(\ref{eqmouv}) and the two boundary conditions respectively give
\begin{eqnarray}
\alpha_k^2 = k^2-\Omega_k^2, \; {\rm and} \, 
\alpha_k \tanh(\alpha_k) = b \;  
\Omega_k^2, \label{di}
\end{eqnarray}
where $\Omega_k \equiv \om_k - M k$ is the frequency in the co-moving frame. At low wave number, the dispersion relation reads
\begin{equation}
\Omega_k^2 = c_S^2(b)  k^2 - k^4 /\Lambda_b^2 + O(k^6),
\end{equation}
where 
\begin{eqnarray}\label{csb}
c_S^2(b) = \frac{1}{1+b}, \; \Lambda_b^2 = \frac{3 \lp 1+b \rp^3}{b^2}.
\end{eqnarray}
One sees the important effect of the boundary condition of \eq{CL2}: the low frequency group velocity with respect to the fluid ($= \partial_k\Omega$) is reduced by the compliant wall. One also sees that the dispersive length $1/\Lambda_{b}$ given by the quartic term vanishes in the limit $b \to 0$.

To obtain flows crossing the effective sound speed, we make $b$ vary with $x$, see Fig.~\ref{fig:bx}. We call $b_1 > b_2$ its asymptotic values, and $d_b$ its typical variation length. We choose the following form for $b(x)$: 
\begin{equation} \label{bprofile}
b(x)=\frac{b_1+b_2}{2}+\frac{b_2-b_1}{2}\tanh \left(\frac{x}{d_b} \right).
\end{equation}
We then adjust the flow speed $M$ to obtain 
\be 
c_S(b_1) = \frac{1}{\sqrt{1+b_1}} < \left\lvert M \right\rvert < \frac{1}{\sqrt{1+b_2}} =c_S(b_2).
\ee
Since the background flow is stationary, we shall work with (complex) stationary waves:
\begin{equation}
\tilde{\phi}_\omega(x,y) = \int_{-\infty}^{+\infty}\frac{\mathrm{d}t}{2\pi} e^{i\omega t}\phi(t,x,y). 
\label{statio}
\end{equation}

{\it Wave equation} --
Since the height $H$ of the duct is much smaller than typical longitudinal wavelengths, we expect that stationary waves obey an effective 1 dimensional equation in $x$, as it is the case in elongated atomic Bose condensates~\cite{Macher:2009nz}, and in flumes~\cite{Schutzhold:2002rf,Coutant:2012mf}. To obtain such a reduction is non trivial as the $x$-dependence of $b(x)$ prevents us from factorizing out a $y$-dependent factor. To proceed, and to make contact with the above Refs., it is useful to exploit the fact that \eq{eqmouv} and \eq{CL2} can be derived from the following action 
\eqa{S &= \frac{1}{2}\int \mathrm{d}t \int \mathrm{d}x \int_0^1\mathrm{d}y \, {\cal L} 
\nn {\cal L} &= ( D_t\phi )^2 - ( \nabla\phi )^2  + \delta(y -1) b(x) (D_t \phi)^2 .\label{Actionsimple}}
We notice that the non-trivial condition of \eq{CL2} is incorporated by the above boundary term. Introducing the conjugate momentum $ \pi(t,x,y) = \partial {\cal L}/\partial (\partial_t\phi) $ one obtains the Hamiltonian H by the usual Legendre transformation. In addition, as in~\cite{Unruh:1994je,Coutant:2012mf}, the conserved inner product $(\cdot \vert \cdot )$ is not positive definite, and has the Klein-Gordon form,
\begin{eqnarray}
(\phi_1 ,\phi_{2}) &=& i\int_{-\infty}^{+\infty}\mathrm{d}x\int_{0}^{1}\mathrm{d}y \left(\pi_{1}^* \phi_{2} - 
\phi_{1}^* \pi_{2}\right), 
\label{ps}
\end{eqnarray}
where $\phi_{1},\phi_{2}$ are two complex solutions of Eq.~(\ref{eqmouv}, \ref{CL2}), and $\pi_{1},\pi_{2}$ their associated momenta. \eq{ps} is conserved in virtue of Hamilton's equations. As for sound waves in other media, it identically vanishes for all real solutions. However it provides a key information when studying stationary modes, namely the sign of their norm $\lp \phi_\om,\phi_\om \rp$. Indeed, as we shall see, for a fixed $\omega > 0$, there will be both positive and negative norm modes. When the flow is stationary, for any complex solution $\phi$, the wave energy is conserved and related to \eq{ps} by 
\be \label{eq:Hvssp}
{\rm H}[2 {\rm Re}(\phi)] = \lp \phi, i \pad_t \phi \rp. 
\ee
Moreover, when the flow is also asymptotically homogeneous, for every asymptotic plane wave $\varphi_k \propto \cosh(\alpha_k y) e^{i \lp k x - \omega_k t \rp}$, the sign of H is that of $\om_k \Omega_k$. This relation will allow us to identify the negative energy waves without ambiguity.
 
We can now proceed following the hydrodynamic treatment of \cite{Coutant:2012mf}. As a first step, it is useful to derive a (1+1)-dimensional equation from which an effective space-time metric can be read out. When the (adimensional) wavelength in the $x$ direction is much larger than $1$, we can assume that $\pad_y^2\phi$ is independent of $y$. As $\pad_y \phi=0$ at $y=0$, we write the field as
\be \label{eq:PhiPsi}
\phi(x,y,t) \approx \Phi(x,t) + y^2 \Psi(x,t).
\ee
Plugging this into the action~\eq{Actionsimple} and varying it with respect to $\Phi$ and $\Psi$, we get two coupled equations. Combining them, we obtain $(\hat{\mathcal{O}}_2 + \hat{\mathcal{O}}_4)\Phi=0$, where $\hat{\mathcal{O}}_n$ is a $n^{th}$-order operator in $\pad_t$ and $\pad_x$. The quadratic term is 
$\hat{\mathcal{O}}_2 = \pad_\mu F^{\mu \nu}(x) \pad_\nu$, where 
\be \label{eq:metric}
F^{\mu \nu}(x) = 
\begin{pmatrix}
c_S^2(b(x)) - M^2 & M \\
M & 1 
\end{pmatrix} ,
\ee
and $c_S^2(b(x))$ is given in \eq{csb}. Up to a conformal factor, we obtain the d'Alembert equation in a two dimensional space-time with metric $g^{\mu \nu} \propto F^{\mu \nu}$. This metric has a Killing horizon where $c_S^2(x) - M^2$ vanishes~\cite{Barcelo:2005fc}. This correspondence with gravity relates the anomalous scattering described below to the Hawking effect. 

Contrary to what happens for sound waves in atomic BECs, or water waves in the incompressible limit, $\hat{\mathcal{O}}_4$ also contains third and fourth derivatives in time. This prevents to apply the standard treatment on the sole field $\Phi$. However, the set of two coupled equations on $(\Phi, \Psi)$ is hamiltonian and can be used to study the scattering. Alternatively, one can work with the original model in $2+1$ dimensions based on \eq{Actionsimple}. We performed numerical simulations with both models and found similar results.

{\it Anomalous mode mixing} --
Since stationary waves with different frequencies $\omega$ do not mix, the scattering only concerns the discrete set of modes with the same $\omega$. To characterize it, we identify its dimensionality and the norms of the various asymptotic modes for $x \to \pm \infty$. 

\begin{figure}[h]
\includegraphics[width=\linewidth]{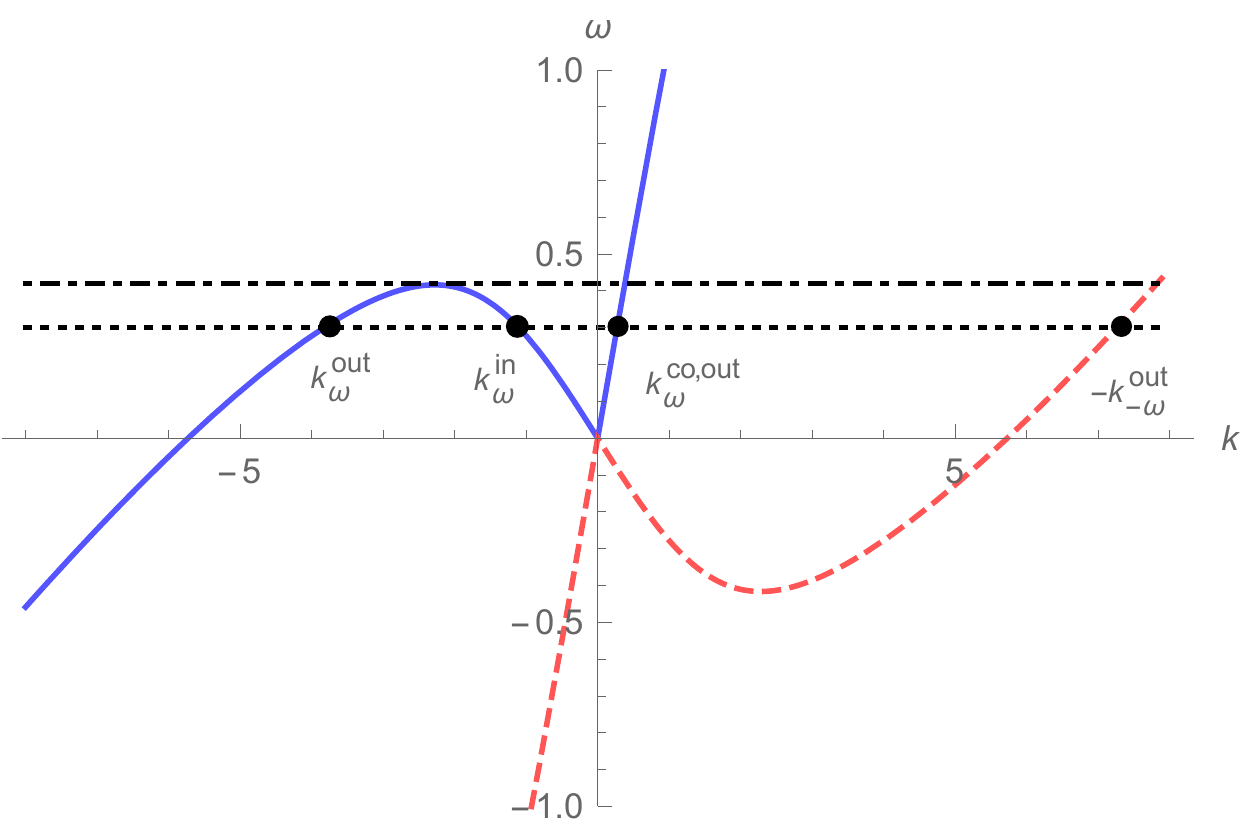} 
\caption{Dispersion relation $\omega$ versus $k$ in a homogeneous subsonic flow. The blue, plain curves show the roots with positive comoving frequency $\Omega$, and the red, dashed ones those with $\Omega < 0$. The dotted black line represents $\om = 0.4$, and the dot-dashed one shows the value $\om_{\rm max}$ of $\om$ at which the two roots with $k< 0$ merge. The parameters are $b=1$, $M=0.4$, and the effective sound velocity $c_S$ is equal to $1/\sqrt{2} \approx 0.71$.}\label{figdispertion}
\end{figure}

\Fig{figdispertion} shows the dispersion relation and the roots at fixed $\om$ in a subsonic flow for $M>0$. For this sign of $M$, the flow associated with \eq{bprofile} passes from supersonic to subsonic along the direction of the stream. It thus corresponds to a white-hole flow, as those studied in~\cite{Weinfurtner:2010nu,Coutant:2012mf,Michel:2014zsa}.~\footnote{For $M<0$, one would describe a black hole flow. The forthcoming analysis applies by reversing the sign of velocities and the ``in'' or ``out'' character of the modes, see~\cite{Macher:2009tw}.} For definiteness, we discuss only the case $\om > 0$. The same results are directly applicable to $\om<0$ after complex conjugation. In the subsonic region, on the right of the horizon, there exists a critical frequency $\om_{\rm max}$ (close to $0.42$ in the Figure) at which two roots merge. For $\om < \om_{\rm max}$, the dispersion relation has 4 real roots.\footnote{In the supersonic region, only two real wave-vectors remain: $k_\om^{\rm co, in}$ and $-k_{-\om}^{\rm in}$. They both describe incoming modes.} Following~\cite{Coutant:2012mf}, we call their wave-vectors $k_\om^{\rm out}$, $k_\om^{\rm in}$, $k_\om^{\rm co, out}$, and $-k_{-\om}^{\rm out}$. The corresponding asymptotic modes are, respectively, $\varphi_\om^{\rm out}$, $\varphi_\om^{\rm in}$, $\varphi_\om^{\rm co, out}$, and $\lp \varphi_{-\om}^{\rm out} \rp^*$. They are characterized by three important properties:

1) in- or out-character: $\varphi_\om^{\rm in}$ is incoming (it moves towards the horizon) while the three other modes are outgoing (they move away from the horizon);

2) energy sign: $\lp \varphi_{-\om}^{\rm out} \rp^*$ carries a negative energy and a negative norm, see~\eq{eq:Hvssp}. (It has been complex conjugated so that $\varphi_{-\om}^{\rm out}$ is a positive norm mode). The three other modes have positive energy and norm;

3) co- or counter-propagating nature: $\varphi_\om^{\rm co, out}$ is co-propagating (its group velocity in the {\it frame of the fluid} is positive) while the three others are counter-propagating. This separation is useful because only the latter are significantly mixed in a transonic flow~\cite{Macher:2009tw,Coutant:2011in}. In effect, $\varphi_\om^{\rm co, out}$ acts essentially as a spectator.

To get the $S$ matrix, we need to identify the basis of incoming (outgoing) modes $\phi_\om^{\rm in}$ ($\phi_\om^{\rm out}$) which contain only one incoming (resp. outgoing) asymptotic plane wave. For $\om < \om_{\rm max}$, there are three modes, so the scattering matrix has a size $3 \times 3$, and is an element of the Lie group $U(2,1)$ since $\lp \varphi_{-\om} \rp^*$ has a negative norm~\cite{Macher:2009tw}. In this paper we focus on the incident mode $\phi_\om^{\rm in}$ for a white hole flow. It is a good candidate to probe the analogue Hawking effect, and was studied in hydrodynamic flows~\cite{Rousseaux:2007is,Weinfurtner:2010nu,Michel:2014zsa,Euve:2014aga}. For ${x \to \infty}$, it is a sum of four asymptotic modes
\eqa{\label{eq:inmode}
\phi_\om^{in} \approx \varphi_\om^{\rm in} + \alpha_\om  \varphi_\om^{\rm out} + \beta_\om  \lp \varphi_{-\om}^{\rm out} \rp^* + A_\om \varphi_\om^{co, \, out}.
} 
In transonic flows, there is no transmitted wave~\cite{Michel:2014zsa}. $\phi_\om^{in}$ thus vanishes for ${x \to - \infty}$. When working with asymptotic modes of unit norm, the norm of $\phi_\om^{in}$ evaluated at late time (in the sense of a broad wave packet) 
\eqa{\label{eq:normrel}
 (N^{\rm out}_\om)^2 = \left\lvert \alpha_\om \right\rvert^2 - \left\lvert \beta_\om \right\rvert^2 + \left\lvert A_\om \right\rvert^2 ,
}
must be exactly 1 because of the conservation of \eq{ps}. (In stationary flows, $(N^{\rm out}_\om)^2 = 1$ also expresses the conservation of the wave energy, see \eq{eq:Hvssp}, and that of the energy flux of M\"ohring~\cite{FLM:68101}.) The minus sign in front of $| \beta_\om |^2 $ is the signature of an anomalous scattering. It stems from the negative norm carried by $\lp \phi_{-\om}^{\rm out} \rp^*$, see the above point 2. The coefficient $\beta_\om$ thus mixes modes of opposite norms and energies. In quantum settings, $\left\lvert \beta_\om \right\rvert^2$ would give the mean number of spontaneously produced particles from amplifying vacuum fluctuations, that is, the Hawking radiation~\cite{Hawking:1974sw}. The gravitational analogy~\cite{Unruh:1980cg,Unruh:1994je} indicates that $\left\lvert \beta_\om \right\rvert^2$ should follow a Planck law when dispersion effects (and grey body factors~\cite{Page:1976df,Anderson:2014jua}) are negligible. Moreover, it predicts that the effective temperature $T$ should be given by $T_H = \kappa/2\pi$, where $\kappa$ is the surface gravity obtained from the analogue metric of \eq{eq:metric}. (We choose the units so that $k_B/\hbar = 1$. $T_H$ and $\kappa$ are thus both frequencies.) Using \eq{csb}, one gets 
\be
\kappa = \partial_x c_S \vert_{c_S = M} = -\frac{M^3}{2} \partial_x b \vert_{c_S = M}. 
\label{kappa}
\ee

\begin{figure}
\includegraphics[width=\linewidth]{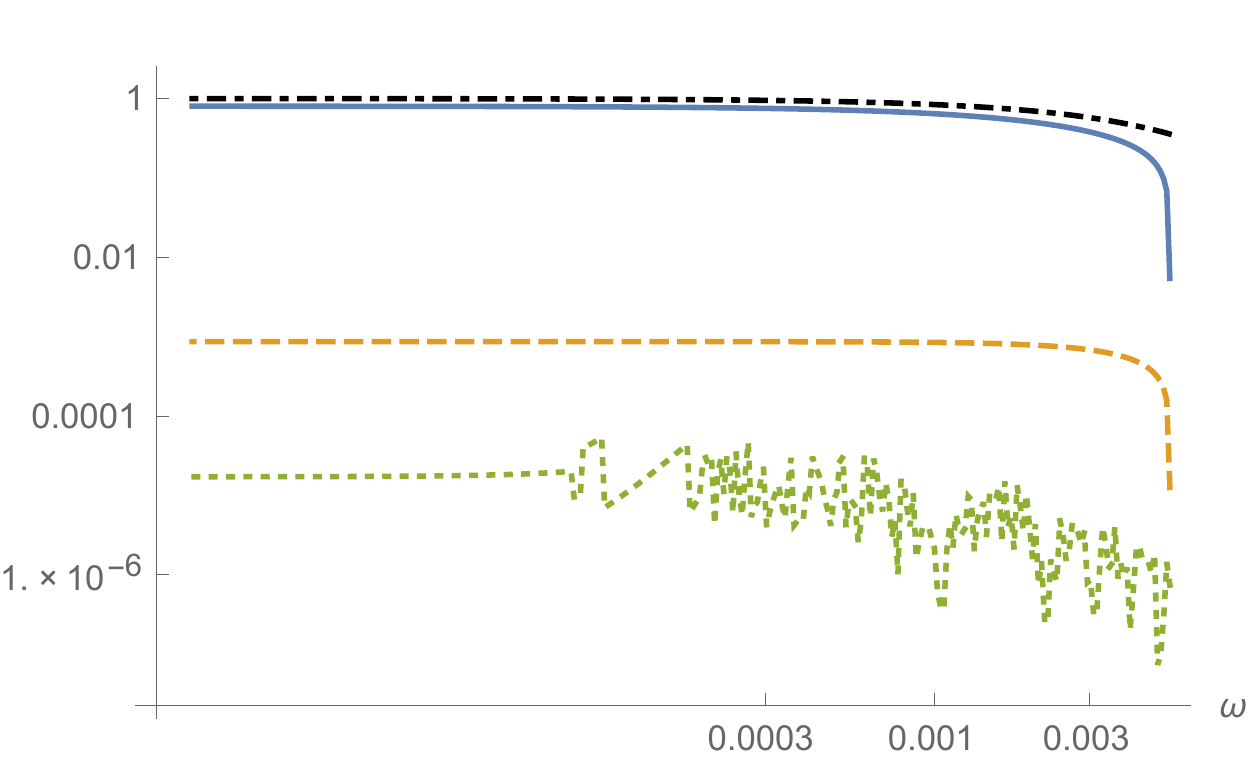}
\caption{Plot of $\frac{\om}{T_H} \left\lvert \beta_\om \right\rvert^2$ (blue, plain) and $\left\lvert A_\om \right\rvert^2$ (orange, dashed) of \eq{eq:inmode} as functions of the frequency. The black, dot-dashed curve shows $\frac{\om}{T_H} \left\lvert \beta_\om \right\rvert^2$ for a Planck law at the Hawking temperature $T_H = \kappa/(2 \pi)$. The parameters are: $M = 1/3, b_1=9, b_2=7$, and $d_b=1$. The green, dotted line represents $(N^{out}_\om)^2 - 1$ where $(N^{out}_\om)^2$ is given in \eq{eq:normrel}.  Its non-vanishing value quantifies the numerical errors.} \label{fig:WH}
\end{figure}

{\it Spectral analysis} --
We numerically solved the set of coupled equations on the $(1+1)$-dimensional fields $\Phi$ and $\Psi$, 
using the method of~\cite{Michel:2014zsa} adapted to the present case. 
The results concerning the incoming mode of \eq{eq:inmode} propagating in a transonic flow described by \eq{bprofile} are shown in~\fig{fig:WH}.~\footnote{When sending a localized wave-packet on a white hole horizon, we observed at late times the formation of an undulation, see~\cite{suppl}.} 
We stopped the integration for $\om$ slightly below the critical frequency $\om_{\rm max}$, where $\beta_\om$ and $A_\om$ both vanish. We tuned the various parameters (given in the caption of \fig{fig:WH}), so that the flow is near-critical: $M/c_S(b_1) \approx 1.054$ and $M/c_S(b_2) \approx 0.943$. Using these parameters, one has $\om_{\rm max}\approx 0.0053$, and $\kappa  \approx 0.019$ of the same order as the dispersive frequency scale $\Lambda_b c_S^2$ evaluated at the horizon. This means that we worked just outside the weak dispersive regime~\cite{Finazzi:2012iu}. Yet, for frequencies up to $\om_{\rm max}$, $| \beta_\om |^2$ follows rather well the Hawking prediction $| \beta^H_\om |^2 =1/( e^{\omega/ T_H}- 1 )$, that is, a Planck law with $T_H$ given by $\kappa / (2 \pi)$, see \eq{kappa}. At low frequency, the relative difference $ (| \beta_\om |^2/| \beta^H_\om |^2 )- 1$ is of order $20 \%$ (when we used $d_b=3$, the difference reduced to about $0.3 \%$, as expected since we were then in a weakly dispersive regime). Moreover, we see that the coefficient $A_\om$ involving the co-propagating mode can be safely neglected as $\left\lvert A_\om \right\rvert^2$ remains smaller that $0.1 \%$.  In order to estimate the numerical errors, we also show the quantity $(N^{out}_\om)^2 - 1$ which must be equal to zero, as explained below \eq{eq:normrel}. In brief, the properties we obtain are in close agreement with those found in other media~\cite{Macher:2009tw,Macher:2009nz,Robertson:2012ku,Michel:2013wpa}.

{\it Experimental aspects} --
The incident waves will be sent by a loudspeaker in a rectangular channel, and the scattered waves will be observed using two arrays of microphones, see \fig{fig:bx}. The frequency range will be chosen to fulfill the low frequency hypothesis used in \eq{CL2}. A stationary flow with a mean Mach number $M \sim 0.3$ will be provided in the channel. 
The compliant wall will be realized with a honeycomb structure. Its height $H\, b(x)$ will vary in such a way that the flow is transcritical, i.e., that $M / c_S$ crosses unity at some $x$, something which has not yet been reached in water tanks experiments~\cite{Rousseaux:2007is,Weinfurtner:2010nu} aiming at detecting the analogue Hawking radiation. Another important advantage is the possibility to send the three types of incident waves, and therefore to measure the 9 scattering coefficients, both in black hole and white hole flows. We suspect that working in these conditions can produce turbulence effects and whistling. To reduce these effects, the compliant wall will be covered by a wire gauze with a very low flow resistance. Special attention should also be devoted to dissipation. In spite of these experimental difficulties, the system seems to be a good candidate to probe the various aspects of the analogue Hawking radiation.

{\it Conclusions} --
We showed that a low Mach number uniform flow of air in a tube with a compliant wall can produce a sonic horizon by reducing the local effective one-dimensional speed of sound. Despite the unusual boundary condition at the compliant wall, the problem was phrased in a Hamiltonian formalism. For near-critical flows, a (1+1) dimensional-reduction was performed, exhibiting an analogue metric and the first dispersion terms while retaining the hamiltonian structure. The Hawking spectrum was numerically recovered for sufficiently slowly varying profiles of the compliant wall. We thus hope to be able to verify that the norm of the anomalous coefficient $|\beta_\omega|^2$ grows as $\kappa /(2\pi \omega)$ for low frequency.

{\it Acknowledgements} --
We thank Scott Robertson for his remarks on a preliminary version of this letter. PF is grateful to the LPT for its hospitality during an internship sponsored by the École Normale supérieure. 

\bibliographystyle{apsrev4-1}
\bibliography{bibliopubli}

\begin{thebibliography}{32}%
\makeatletter
\providecommand \@ifxundefined [1]{%
 \@ifx{#1\undefined}
}%
\providecommand \@ifnum [1]{%
 \ifnum #1\expandafter \@firstoftwo
 \else \expandafter \@secondoftwo
 \fi
}%
\providecommand \@ifx [1]{%
 \ifx #1\expandafter \@firstoftwo
 \else \expandafter \@secondoftwo
 \fi
}%
\providecommand \natexlab [1]{#1}%
\providecommand \enquote  [1]{``#1''}%
\providecommand \bibnamefont  [1]{#1}%
\providecommand \bibfnamefont [1]{#1}%
\providecommand \citenamefont [1]{#1}%
\providecommand \href@noop [0]{\@secondoftwo}%
\providecommand \href [0]{\begingroup \@sanitize@url \@href}%
\providecommand \@href[1]{\@@startlink{#1}\@@href}%
\providecommand \@@href[1]{\endgroup#1\@@endlink}%
\providecommand \@sanitize@url [0]{\catcode `\\12\catcode `\$12\catcode
  `\&12\catcode `\#12\catcode `\^12\catcode `\_12\catcode `\%12\relax}%
\providecommand \@@startlink[1]{}%
\providecommand \@@endlink[0]{}%
\providecommand \url  [0]{\begingroup\@sanitize@url \@url }%
\providecommand \@url [1]{\endgroup\@href {#1}{\urlprefix }}%
\providecommand \urlprefix  [0]{URL }%
\providecommand \Eprint [0]{\href }%
\providecommand \doibase [0]{http://dx.doi.org/}%
\providecommand \selectlanguage [0]{\@gobble}%
\providecommand \bibinfo  [0]{\@secondoftwo}%
\providecommand \bibfield  [0]{\@secondoftwo}%
\providecommand \translation [1]{[#1]}%
\providecommand \BibitemOpen [0]{}%
\providecommand \bibitemStop [0]{}%
\providecommand \bibitemNoStop [0]{.\EOS\space}%
\providecommand \EOS [0]{\spacefactor3000\relax}%
\providecommand \BibitemShut  [1]{\csname bibitem#1\endcsname}%
\let\auto@bib@innerbib\@empty
\bibitem [{\citenamefont {Unruh}(1981)}]{Unruh:1980cg}%
  \BibitemOpen
  \bibfield  {author} {\bibinfo {author} {\bibfnamefont {W.}~\bibnamefont
  {Unruh}},\ }\href {\doibase 10.1103/PhysRevLett.46.1351} {\bibfield
  {journal} {\bibinfo  {journal} {Phys. Rev. Lett.}\ }\textbf {\bibinfo
  {volume} {46}},\ \bibinfo {pages} {1351} (\bibinfo {year}
  {1981})}\BibitemShut {NoStop}%
\bibitem [{\citenamefont {Hawking}(1975)}]{Hawking:1974sw}%
  \BibitemOpen
  \bibfield  {author} {\bibinfo {author} {\bibfnamefont {S.~W.}\ \bibnamefont
  {Hawking}},\ }\href {\doibase 10.1007/BF02345020} {\bibfield  {journal}
  {\bibinfo  {journal} {Commun. Math. Phys.}\ }\textbf {\bibinfo {volume}
  {43}},\ \bibinfo {pages} {199} (\bibinfo {year} {1975})}\BibitemShut
  {NoStop}%
\bibitem [{\citenamefont {Brout}\ \emph
  {et~al.}(1995{\natexlab{a}})\citenamefont {Brout}, \citenamefont {Massar},
  \citenamefont {Parentani},\ and\ \citenamefont {Spindel}}]{Brout:1995rd}%
  \BibitemOpen
  \bibfield  {author} {\bibinfo {author} {\bibfnamefont {R.}~\bibnamefont
  {Brout}}, \bibinfo {author} {\bibfnamefont {S.}~\bibnamefont {Massar}},
  \bibinfo {author} {\bibfnamefont {R.}~\bibnamefont {Parentani}}, \ and\
  \bibinfo {author} {\bibfnamefont {P.}~\bibnamefont {Spindel}},\ }\href
  {\doibase 10.1016/0370-1573(95)00008-5} {\bibfield  {journal} {\bibinfo
  {journal} {Phys. Rept.}\ }\textbf {\bibinfo {volume} {260}},\ \bibinfo
  {pages} {329} (\bibinfo {year} {1995}{\natexlab{a}})},\ \Eprint
  {http://arxiv.org/abs/0710.4345} {arXiv:0710.4345 [gr-qc]} \BibitemShut
  {NoStop}%
\bibitem [{\citenamefont {Unruh}(1995)}]{Unruh:1994je}%
  \BibitemOpen
  \bibfield  {author} {\bibinfo {author} {\bibfnamefont {W.}~\bibnamefont
  {Unruh}},\ }\href {\doibase 10.1103/PhysRevD.51.2827} {\bibfield  {journal}
  {\bibinfo  {journal} {Phys. Rev. D}\ }\textbf {\bibinfo {volume} {51}},\
  \bibinfo {pages} {2827} (\bibinfo {year} {1995})}\BibitemShut {NoStop}%
\bibitem [{\citenamefont {Brout}\ \emph
  {et~al.}(1995{\natexlab{b}})\citenamefont {Brout}, \citenamefont {Massar},
  \citenamefont {Parentani},\ and\ \citenamefont {Spindel}}]{Brout:1995wp}%
  \BibitemOpen
  \bibfield  {author} {\bibinfo {author} {\bibfnamefont {R.}~\bibnamefont
  {Brout}}, \bibinfo {author} {\bibfnamefont {S.}~\bibnamefont {Massar}},
  \bibinfo {author} {\bibfnamefont {R.}~\bibnamefont {Parentani}}, \ and\
  \bibinfo {author} {\bibfnamefont {P.}~\bibnamefont {Spindel}},\ }\href
  {\doibase 10.1103/PhysRevD.52.4559} {\bibfield  {journal} {\bibinfo
  {journal} {Phys. Rev. D}\ }\textbf {\bibinfo {volume} {52}},\ \bibinfo
  {pages} {4559} (\bibinfo {year} {1995}{\natexlab{b}})},\ \Eprint
  {http://arxiv.org/abs/hep-th/9506121} {arXiv:hep-th/9506121 [hep-th]}
  \BibitemShut {NoStop}%
\bibitem [{\citenamefont {Horava}(2009)}]{Horava:2009uw}%
  \BibitemOpen
  \bibfield  {author} {\bibinfo {author} {\bibfnamefont {P.}~\bibnamefont
  {Horava}},\ }\href {\doibase 10.1103/PhysRevD.79.084008} {\bibfield
  {journal} {\bibinfo  {journal} {Phys. Rev. D}\ }\textbf {\bibinfo {volume}
  {79}},\ \bibinfo {pages} {084008} (\bibinfo {year} {2009})},\ \Eprint
  {http://arxiv.org/abs/0901.3775} {arXiv:0901.3775 [hep-th]} \BibitemShut
  {NoStop}%
\bibitem [{\citenamefont {Jacobson}(2010)}]{Jacobson:2010mx}%
  \BibitemOpen
  \bibfield  {author} {\bibinfo {author} {\bibfnamefont {T.}~\bibnamefont
  {Jacobson}},\ }\href {\doibase 10.1103/PhysRevD.82.129901,
  10.1103/PhysRevD.81.101502} {\bibfield  {journal} {\bibinfo  {journal}
  {Phys.Rev.}\ }\textbf {\bibinfo {volume} {D81}},\ \bibinfo {pages} {101502}
  (\bibinfo {year} {2010})},\ \Eprint {http://arxiv.org/abs/1001.4823}
  {arXiv:1001.4823 [hep-th]} \BibitemShut {NoStop}%
\bibitem [{\citenamefont {Coutant}\ and\ \citenamefont
  {Parentani}(2014{\natexlab{a}})}]{Coutant:2014cwa}%
  \BibitemOpen
  \bibfield  {author} {\bibinfo {author} {\bibfnamefont {A.}~\bibnamefont
  {Coutant}}\ and\ \bibinfo {author} {\bibfnamefont {R.}~\bibnamefont
  {Parentani}},\ }\href {\doibase 10.1103/PhysRevD.90.121501} {\bibfield
  {journal} {\bibinfo  {journal} {Phys.Rev.}\ }\textbf {\bibinfo {volume}
  {D90}},\ \bibinfo {pages} {121501} (\bibinfo {year} {2014}{\natexlab{a}})},\
  \Eprint {http://arxiv.org/abs/1402.2514} {arXiv:1402.2514 [gr-qc]}
  \BibitemShut {NoStop}%
\bibitem [{\citenamefont {Macher}\ and\ \citenamefont
  {Parentani}(2009{\natexlab{a}})}]{Macher:2009tw}%
  \BibitemOpen
  \bibfield  {author} {\bibinfo {author} {\bibfnamefont {J.}~\bibnamefont
  {Macher}}\ and\ \bibinfo {author} {\bibfnamefont {R.}~\bibnamefont
  {Parentani}},\ }\href {\doibase 10.1103/PhysRevD.79.124008} {\bibfield
  {journal} {\bibinfo  {journal} {Phys. Rev. D}\ }\textbf {\bibinfo {volume}
  {79}},\ \bibinfo {pages} {124008} (\bibinfo {year} {2009}{\natexlab{a}})},\
  \Eprint {http://arxiv.org/abs/0903.2224} {arXiv:0903.2224 [hep-th]}
  \BibitemShut {NoStop}%
\bibitem [{\citenamefont {Coutant}\ \emph {et~al.}(2012)\citenamefont
  {Coutant}, \citenamefont {Parentani},\ and\ \citenamefont
  {Finazzi}}]{Coutant:2011in}%
  \BibitemOpen
  \bibfield  {author} {\bibinfo {author} {\bibfnamefont {A.}~\bibnamefont
  {Coutant}}, \bibinfo {author} {\bibfnamefont {R.}~\bibnamefont {Parentani}},
  \ and\ \bibinfo {author} {\bibfnamefont {S.}~\bibnamefont {Finazzi}},\ }\href
  {\doibase 10.1103/PhysRevD.85.024021} {\bibfield  {journal} {\bibinfo
  {journal} {Phys. Rev. D}\ }\textbf {\bibinfo {volume} {85}},\ \bibinfo
  {pages} {024021} (\bibinfo {year} {2012})},\ \Eprint
  {http://arxiv.org/abs/1108.1821} {arXiv:1108.1821 [hep-th]} \BibitemShut
  {NoStop}%
\bibitem [{\citenamefont {Finazzi}\ and\ \citenamefont
  {Parentani}(2012)}]{Finazzi:2012iu}%
  \BibitemOpen
  \bibfield  {author} {\bibinfo {author} {\bibfnamefont {S.}~\bibnamefont
  {Finazzi}}\ and\ \bibinfo {author} {\bibfnamefont {R.}~\bibnamefont
  {Parentani}},\ }\href {\doibase 10.1103/PhysRevD.85.124027} {\bibfield
  {journal} {\bibinfo  {journal} {Phys. Rev. D}\ }\textbf {\bibinfo {volume}
  {85}},\ \bibinfo {pages} {124027} (\bibinfo {year} {2012})},\ \Eprint
  {http://arxiv.org/abs/1202.6015} {arXiv:1202.6015 [gr-qc]} \BibitemShut
  {NoStop}%
\bibitem [{\citenamefont {Robertson}(2012)}]{Robertson:2012ku}%
  \BibitemOpen
  \bibfield  {author} {\bibinfo {author} {\bibfnamefont {S.~J.}\ \bibnamefont
  {Robertson}},\ }\href {\doibase 10.1088/0953-4075/45/16/163001} {\bibfield
  {journal} {\bibinfo  {journal} {J. Phys. B}\ }\textbf {\bibinfo {volume}
  {45}},\ \bibinfo {pages} {163001} (\bibinfo {year} {2012})}\BibitemShut
  {NoStop}%
\bibitem [{\citenamefont {Garay}\ \emph {et~al.}(2000)\citenamefont {Garay},
  \citenamefont {Anglin}, \citenamefont {Cirac},\ and\ \citenamefont
  {Zoller}}]{Garay:1999sk}%
  \BibitemOpen
  \bibfield  {author} {\bibinfo {author} {\bibfnamefont {L.}~\bibnamefont
  {Garay}}, \bibinfo {author} {\bibfnamefont {J.}~\bibnamefont {Anglin}},
  \bibinfo {author} {\bibfnamefont {J.}~\bibnamefont {Cirac}}, \ and\ \bibinfo
  {author} {\bibfnamefont {P.}~\bibnamefont {Zoller}},\ }\href {\doibase
  10.1103/PhysRevLett.85.4643} {\bibfield  {journal} {\bibinfo  {journal}
  {Phys.Rev.Lett.}\ }\textbf {\bibinfo {volume} {85}},\ \bibinfo {pages} {4643}
  (\bibinfo {year} {2000})},\ \Eprint {http://arxiv.org/abs/gr-qc/0002015}
  {arXiv:gr-qc/0002015 [gr-qc]} \BibitemShut {NoStop}%
\bibitem [{\citenamefont {Schutzhold}\ and\ \citenamefont
  {Unruh}(2002)}]{Schutzhold:2002rf}%
  \BibitemOpen
  \bibfield  {author} {\bibinfo {author} {\bibfnamefont {R.}~\bibnamefont
  {Schutzhold}}\ and\ \bibinfo {author} {\bibfnamefont {W.~G.}\ \bibnamefont
  {Unruh}},\ }\href {\doibase 10.1103/PhysRevD.66.044019} {\bibfield  {journal}
  {\bibinfo  {journal} {Phys. Rev. D}\ }\textbf {\bibinfo {volume} {66}},\
  \bibinfo {pages} {044019} (\bibinfo {year} {2002})},\ \Eprint
  {http://arxiv.org/abs/gr-qc/0205099} {arXiv:gr-qc/0205099 [gr-qc]}
  \BibitemShut {NoStop}%
\bibitem [{\citenamefont {Philbin}\ \emph {et~al.}(2008)\citenamefont
  {Philbin}, \citenamefont {Kuklewicz}, \citenamefont {Robertson},
  \citenamefont {Hill}, \citenamefont {König},\ and\ \citenamefont
  {Leonhardt}}]{Philbin07032008}%
  \BibitemOpen
  \bibfield  {author} {\bibinfo {author} {\bibfnamefont {T.~G.}\ \bibnamefont
  {Philbin}}, \bibinfo {author} {\bibfnamefont {C.}~\bibnamefont {Kuklewicz}},
  \bibinfo {author} {\bibfnamefont {S.}~\bibnamefont {Robertson}}, \bibinfo
  {author} {\bibfnamefont {S.}~\bibnamefont {Hill}}, \bibinfo {author}
  {\bibfnamefont {F.}~\bibnamefont {König}}, \ and\ \bibinfo {author}
  {\bibfnamefont {U.}~\bibnamefont {Leonhardt}},\ }\href {\doibase
  10.1126/science.1153625} {\bibfield  {journal} {\bibinfo  {journal}
  {Science}\ }\textbf {\bibinfo {volume} {319}},\ \bibinfo {pages} {1367}
  (\bibinfo {year} {2008})},\ \Eprint
  {http://arxiv.org/abs/http://www.sciencemag.org/content/319/5868/1367.full.pdf}
  {http://www.sciencemag.org/content/319/5868/1367.full.pdf} \BibitemShut
  {NoStop}%
\bibitem [{\citenamefont {Rousseaux}\ \emph {et~al.}(2008)\citenamefont
  {Rousseaux}, \citenamefont {Mathis}, \citenamefont {Maissa}, \citenamefont
  {Philbin},\ and\ \citenamefont {Leonhardt}}]{Rousseaux:2007is}%
  \BibitemOpen
  \bibfield  {author} {\bibinfo {author} {\bibfnamefont {G.}~\bibnamefont
  {Rousseaux}}, \bibinfo {author} {\bibfnamefont {C.}~\bibnamefont {Mathis}},
  \bibinfo {author} {\bibfnamefont {P.}~\bibnamefont {Maissa}}, \bibinfo
  {author} {\bibfnamefont {T.~G.}\ \bibnamefont {Philbin}}, \ and\ \bibinfo
  {author} {\bibfnamefont {U.}~\bibnamefont {Leonhardt}},\ }\href {\doibase
  10.1088/1367-2630/10/5/053015} {\bibfield  {journal} {\bibinfo  {journal}
  {New J. Phys.}\ }\textbf {\bibinfo {volume} {10}},\ \bibinfo {pages} {053015}
  (\bibinfo {year} {2008})},\ \Eprint {http://arxiv.org/abs/0711.4767}
  {arXiv:0711.4767 [gr-qc]} \BibitemShut {NoStop}%
\bibitem [{\citenamefont {Weinfurtner}\ \emph {et~al.}(2011)\citenamefont
  {Weinfurtner}, \citenamefont {Tedford}, \citenamefont {Penrice},
  \citenamefont {Unruh},\ and\ \citenamefont {Lawrence}}]{Weinfurtner:2010nu}%
  \BibitemOpen
  \bibfield  {author} {\bibinfo {author} {\bibfnamefont {S.}~\bibnamefont
  {Weinfurtner}}, \bibinfo {author} {\bibfnamefont {E.~W.}\ \bibnamefont
  {Tedford}}, \bibinfo {author} {\bibfnamefont {M.~C.}\ \bibnamefont
  {Penrice}}, \bibinfo {author} {\bibfnamefont {W.~G.}\ \bibnamefont {Unruh}},
  \ and\ \bibinfo {author} {\bibfnamefont {G.~A.}\ \bibnamefont {Lawrence}},\
  }\href {\doibase 10.1103/PhysRevLett.106.021302} {\bibfield  {journal}
  {\bibinfo  {journal} {Phys. Rev. Lett.}\ }\textbf {\bibinfo {volume} {106}},\
  \bibinfo {pages} {021302} (\bibinfo {year} {2011})},\ \Eprint
  {http://arxiv.org/abs/1008.1911} {arXiv:1008.1911 [gr-qc]} \BibitemShut
  {NoStop}%
\bibitem [{\citenamefont {{Lahav}}\ \emph {et~al.}(2010)\citenamefont
  {{Lahav}}, \citenamefont {{Itah}}, \citenamefont {{Blumkin}}, \citenamefont
  {{Gordon}}, \citenamefont {{Rinott}}, \citenamefont {{Zayats}},\ and\
  \citenamefont {{Steinhauer}}}]{2010PhRvL.105x0401L}%
  \BibitemOpen
  \bibfield  {author} {\bibinfo {author} {\bibfnamefont {O.}~\bibnamefont
  {{Lahav}}}, \bibinfo {author} {\bibfnamefont {A.}~\bibnamefont {{Itah}}},
  \bibinfo {author} {\bibfnamefont {A.}~\bibnamefont {{Blumkin}}}, \bibinfo
  {author} {\bibfnamefont {C.}~\bibnamefont {{Gordon}}}, \bibinfo {author}
  {\bibfnamefont {S.}~\bibnamefont {{Rinott}}}, \bibinfo {author}
  {\bibfnamefont {A.}~\bibnamefont {{Zayats}}}, \ and\ \bibinfo {author}
  {\bibfnamefont {J.}~\bibnamefont {{Steinhauer}}},\ }\href {\doibase
  10.1103/PhysRevLett.105.240401} {\bibfield  {journal} {\bibinfo  {journal}
  {Physical Review Letters}\ }\textbf {\bibinfo {volume} {105}},\ \bibinfo
  {eid} {240401} (\bibinfo {year} {2010})},\ \Eprint
  {http://arxiv.org/abs/0906.1337} {arXiv:0906.1337 [cond-mat.quant-gas]}
  \BibitemShut {NoStop}%
\bibitem [{\citenamefont {{Steinhauer}}(2014)}]{BHLaser-Jeff}%
  \BibitemOpen
  \bibfield  {author} {\bibinfo {author} {\bibfnamefont {J.}~\bibnamefont
  {{Steinhauer}}},\ }\href {\doibase 10.1038/nphys3104} {\bibfield  {journal}
  {\bibinfo  {journal} {Nature Physics}\ }\textbf {\bibinfo {volume} {10}},\
  \bibinfo {pages} {864} (\bibinfo {year} {2014})},\ \Eprint
  {http://arxiv.org/abs/1409.6550} {arXiv:1409.6550 [cond-mat.quant-gas]}
  \BibitemShut {NoStop}%
\bibitem [{\citenamefont {{Myers}}(1980)}]{1980JSV....71..429M}%
  \BibitemOpen
  \bibfield  {author} {\bibinfo {author} {\bibfnamefont {M.~K.}\ \bibnamefont
  {{Myers}}},\ }\href {\doibase 10.1016/0022-460X(80)90424-1} {\bibfield
  {journal} {\bibinfo  {journal} {Journal of Sound Vibration}\ }\textbf
  {\bibinfo {volume} {71}},\ \bibinfo {pages} {429} (\bibinfo {year}
  {1980})}\BibitemShut {NoStop}%
\bibitem [{\citenamefont {{Auregan}}\ and\ \citenamefont
  {{Pagneux}}(2015)}]{JASA}%
  \BibitemOpen
  \bibfield  {author} {\bibinfo {author} {\bibfnamefont {Y.}~\bibnamefont
  {{Auregan}}}\ and\ \bibinfo {author} {\bibfnamefont {V.}~\bibnamefont
  {{Pagneux}}},\ }\href@noop {} {\  (\bibinfo {year} {2015})},\ \Eprint
  {http://arxiv.org/abs/1502.07883} {arXiv:1502.07883 [physics.class-ph]}
  \BibitemShut {NoStop}%
\bibitem [{sup()}]{suppl}%
  \BibitemOpen
  \href@noop {} {}\bibinfo {note} {See supplementary material
  below}\BibitemShut {NoStop}%
\bibitem [{\citenamefont {Macher}\ and\ \citenamefont
  {Parentani}(2009{\natexlab{b}})}]{Macher:2009nz}%
  \BibitemOpen
  \bibfield  {author} {\bibinfo {author} {\bibfnamefont {J.}~\bibnamefont
  {Macher}}\ and\ \bibinfo {author} {\bibfnamefont {R.}~\bibnamefont
  {Parentani}},\ }\href {\doibase 10.1103/PhysRevA.80.043601} {\bibfield
  {journal} {\bibinfo  {journal} {Phys. Rev. A}\ }\textbf {\bibinfo {volume}
  {80}},\ \bibinfo {pages} {043601} (\bibinfo {year} {2009}{\natexlab{b}})},\
  \Eprint {http://arxiv.org/abs/0905.3634} {arXiv:0905.3634
  [cond-mat.quant-gas]} \BibitemShut {NoStop}%
\bibitem [{\citenamefont {Coutant}\ and\ \citenamefont
  {Parentani}(2014{\natexlab{b}})}]{Coutant:2012mf}%
  \BibitemOpen
  \bibfield  {author} {\bibinfo {author} {\bibfnamefont {A.}~\bibnamefont
  {Coutant}}\ and\ \bibinfo {author} {\bibfnamefont {R.}~\bibnamefont
  {Parentani}},\ }\href {\doibase 10.1063/1.4872025} {\bibfield  {journal}
  {\bibinfo  {journal} {Phys.Fluids}\ }\textbf {\bibinfo {volume} {26}},\
  \bibinfo {pages} {044106} (\bibinfo {year} {2014}{\natexlab{b}})}\BibitemShut
  {NoStop}%
\bibitem [{\citenamefont {Barcelo}\ \emph {et~al.}(2011)\citenamefont
  {Barcelo}, \citenamefont {Liberati},\ and\ \citenamefont
  {Visser}}]{Barcelo:2005fc}%
  \BibitemOpen
  \bibfield  {author} {\bibinfo {author} {\bibfnamefont {C.}~\bibnamefont
  {Barcelo}}, \bibinfo {author} {\bibfnamefont {S.}~\bibnamefont {Liberati}}, \
  and\ \bibinfo {author} {\bibfnamefont {M.}~\bibnamefont {Visser}},\ }\href
  {\doibase 10.12942/lrr-2011-3} {\bibfield  {journal} {\bibinfo  {journal}
  {Living Rev. Rel.}\ }\textbf {\bibinfo {volume} {14}},\ \bibinfo {pages} {3}
  (\bibinfo {year} {2011})},\ \Eprint {http://arxiv.org/abs/gr-qc/0505065}
  {arXiv:gr-qc/0505065 [gr-qc]} \BibitemShut {NoStop}%
\bibitem [{\citenamefont {Michel}\ and\ \citenamefont
  {Parentani}(2014)}]{Michel:2014zsa}%
  \BibitemOpen
  \bibfield  {author} {\bibinfo {author} {\bibfnamefont {F.}~\bibnamefont
  {Michel}}\ and\ \bibinfo {author} {\bibfnamefont {R.}~\bibnamefont
  {Parentani}},\ }\href {\doibase 10.1103/PhysRevD.90.044033} {\bibfield
  {journal} {\bibinfo  {journal} {Phys.Rev.}\ }\textbf {\bibinfo {volume}
  {D90}},\ \bibinfo {pages} {044033} (\bibinfo {year} {2014})},\ \Eprint
  {http://arxiv.org/abs/1404.7482} {arXiv:1404.7482 [gr-qc]} \BibitemShut
  {NoStop}%
\bibitem [{\citenamefont {Euvé}\ \emph {et~al.}(2015)\citenamefont {Euvé},
  \citenamefont {Michel}, \citenamefont {Parentani},\ and\ \citenamefont
  {Rousseaux}}]{Euve:2014aga}%
  \BibitemOpen
  \bibfield  {author} {\bibinfo {author} {\bibfnamefont {L.-P.}\ \bibnamefont
  {Euvé}}, \bibinfo {author} {\bibfnamefont {F.}~\bibnamefont {Michel}},
  \bibinfo {author} {\bibfnamefont {R.}~\bibnamefont {Parentani}}, \ and\
  \bibinfo {author} {\bibfnamefont {G.}~\bibnamefont {Rousseaux}},\ }\href
  {\doibase 10.1103/PhysRevD.91.024020} {\bibfield  {journal} {\bibinfo
  {journal} {Phys.Rev.}\ }\textbf {\bibinfo {volume} {D91}},\ \bibinfo {pages}
  {024020} (\bibinfo {year} {2015})},\ \Eprint {http://arxiv.org/abs/1409.3830}
  {arXiv:1409.3830 [gr-qc]} \BibitemShut {NoStop}%
\bibitem [{\citenamefont {MÖHRING}(2001)}]{FLM:68101}%
  \BibitemOpen
  \bibfield  {author} {\bibinfo {author} {\bibfnamefont {W.}~\bibnamefont
  {MÖHRING}},\ }\href {\doibase 10.1017/S0022112000003050} {\bibfield
  {journal} {\bibinfo  {journal} {Journal of Fluid Mechanics}\ }\textbf
  {\bibinfo {volume} {431}},\ \bibinfo {pages} {223} (\bibinfo {year}
  {2001})}\BibitemShut {NoStop}%
\bibitem [{\citenamefont {Page}(1976)}]{Page:1976df}%
  \BibitemOpen
  \bibfield  {author} {\bibinfo {author} {\bibfnamefont {D.~N.}\ \bibnamefont
  {Page}},\ }\href {\doibase 10.1103/PhysRevD.13.198} {\bibfield  {journal}
  {\bibinfo  {journal} {Phys.Rev.}\ }\textbf {\bibinfo {volume} {D13}},\
  \bibinfo {pages} {198} (\bibinfo {year} {1976})}\BibitemShut {NoStop}%
\bibitem [{\citenamefont {Anderson}\ \emph {et~al.}(2014)\citenamefont
  {Anderson}, \citenamefont {Balbinot}, \citenamefont {Fabbri},\ and\
  \citenamefont {Parentani}}]{Anderson:2014jua}%
  \BibitemOpen
  \bibfield  {author} {\bibinfo {author} {\bibfnamefont {P.~R.}\ \bibnamefont
  {Anderson}}, \bibinfo {author} {\bibfnamefont {R.}~\bibnamefont {Balbinot}},
  \bibinfo {author} {\bibfnamefont {A.}~\bibnamefont {Fabbri}}, \ and\ \bibinfo
  {author} {\bibfnamefont {R.}~\bibnamefont {Parentani}},\ }\href {\doibase
  10.1103/PhysRevD.90.104044} {\bibfield  {journal} {\bibinfo  {journal}
  {Phys.Rev.}\ }\textbf {\bibinfo {volume} {D90}},\ \bibinfo {pages} {104044}
  (\bibinfo {year} {2014})},\ \Eprint {http://arxiv.org/abs/1404.3224}
  {arXiv:1404.3224 [gr-qc]} \BibitemShut {NoStop}%
\bibitem [{\citenamefont {Michel}\ and\ \citenamefont
  {Parentani}(2013)}]{Michel:2013wpa}%
  \BibitemOpen
  \bibfield  {author} {\bibinfo {author} {\bibfnamefont {F.}~\bibnamefont
  {Michel}}\ and\ \bibinfo {author} {\bibfnamefont {R.}~\bibnamefont
  {Parentani}},\ }\href {\doibase 10.1103/PhysRevD.88.125012} {\bibfield
  {journal} {\bibinfo  {journal} {Phys.Rev.}\ }\textbf {\bibinfo {volume}
  {D88}},\ \bibinfo {pages} {125012} (\bibinfo {year} {2013})},\ \Eprint
  {http://arxiv.org/abs/1309.5869} {arXiv:1309.5869 [cond-mat.quant-gas]}
  \BibitemShut {NoStop}%
\bibitem [{\citenamefont {Busch}\ \emph {et~al.}(2014)\citenamefont {Busch},
  \citenamefont {Michel},\ and\ \citenamefont {Parentani}}]{Busch:2014hla}%
  \BibitemOpen
  \bibfield  {author} {\bibinfo {author} {\bibfnamefont {X.}~\bibnamefont
  {Busch}}, \bibinfo {author} {\bibfnamefont {F.}~\bibnamefont {Michel}}, \
  and\ \bibinfo {author} {\bibfnamefont {R.}~\bibnamefont {Parentani}},\ }\href
  {\doibase 10.1103/PhysRevD.90.105005} {\bibfield  {journal} {\bibinfo
  {journal} {Phys.Rev.}\ }\textbf {\bibinfo {volume} {D90}},\ \bibinfo {pages}
  {105005} (\bibinfo {year} {2014})},\ \Eprint {http://arxiv.org/abs/1408.2442}
  {arXiv:1408.2442 [gr-qc]} \BibitemShut {NoStop}%
\end{thebibliography}%

\widetext

\clearpage

\begin{center}
\textbf{\large Supplementary Materials: \\ 
Slow Sound in a duct, effective transonic flows and analogue black holes}
\end{center}
\setcounter{equation}{0}
\setcounter{figure}{0}
\setcounter{table}{0}
\setcounter{page}{1}
\makeatletter
\renewcommand{\theequation}{S\arabic{equation}}
\renewcommand{\thefigure}{S\arabic{figure}}
\renewcommand{\bibnumfmt}[1]{[S#1]}

{\it Wave equation and small-frequency approximation} 

\noindent
When considering a stationary mode $\phi_\om \propto e^{-i \om t}$, the boundary condition at the compliant wall $y=1$ reads~\cite{JASA}
\be 
\pad_y \phi_\om = - \lp -i \om + M \pad_x \rp \lp \frac{\tan \lp \om b(x) \rp }{\om} \lp -i \om + M \pad_x \rp  \rp \phi_\om.
\ee
To avoid having to deal with time derivatives of high orders, we consider a small-frequency limit $\tan \lp \om b \rp / \om \approx b$, to obtain~\eq{CL2}. It must be noted that this simplification does not affect the quadratic term in the dispersion relation, but changes the quartic term. The relative difference is of order $\lp 1- M/c_S \rp^2$ when considering the counter-propagating modes. This approximation is thus valid for all the spanned range of $\om \in [0,\om_{\rm max}]$ provided $M/c_S(b(x))$ remains everywhere close to unity.

{\it Wave equation and normalization in the (1+1)D model} 

\noindent
The two operators $\hat{\mathcal{O}}_2$ and $\hat{\mathcal{O}}_4$ appearing below \eq{eq:PhiPsi} are given by
\eqa{
\hat{\mathcal{O}}_2 &= D_t \lp 1+b(x) \rp D_t - \pad_x^2, \nn
\hat{\mathcal{O}}_4 &= \frac{2}{30} \lp D_t^2 - \pad_x^2 \rp \lp D_t \lp 1+6b(x) \rp D_t - \pad_x^2 \rp.
} 
Asymptotic modes are given by two-component plane waves:
\eqa{
\Phi(x,t) &= u \, e^{i \lp k x - \om t \rp}, \nn
\Psi(x,t) &= v \, e^{i \lp k x - \om t \rp},
}
where $\Phi$, $\Psi$ are defined in \eq{eq:PhiPsi}. The prefactors  $u$ and $v$ are given by 
\eqa{
\frac{u}{v} &= -\frac{1}{3} \frac{\lp 1+3b \rp \Omega^2 - k^2}{\lp 1+b \rp \Omega^2 - k^2}, \\
\lp \lp 1+b \rp M \Omega + k \rp \left\lvert u \right\rvert^2 &+ \frac{2}{3} \lp \lp 1+3b \rp M \Omega + k \rp Re \lp u v^* \rp + \frac{1}{5} \lp \lp 1+5b \rp M \Omega + k \rp \left\lvert v \right\rvert^2 = \pm 1.
}
The first condition comes from the wave equation derived by varying the action \eq{Actionsimple} with respect to $\Phi$, while the second one ensures that all the asymptotic modes have a unit norm (up to a minus sign).

{\it Wave-packet and undulation}

\begin{figure} 
\includegraphics[width=\linewidth]{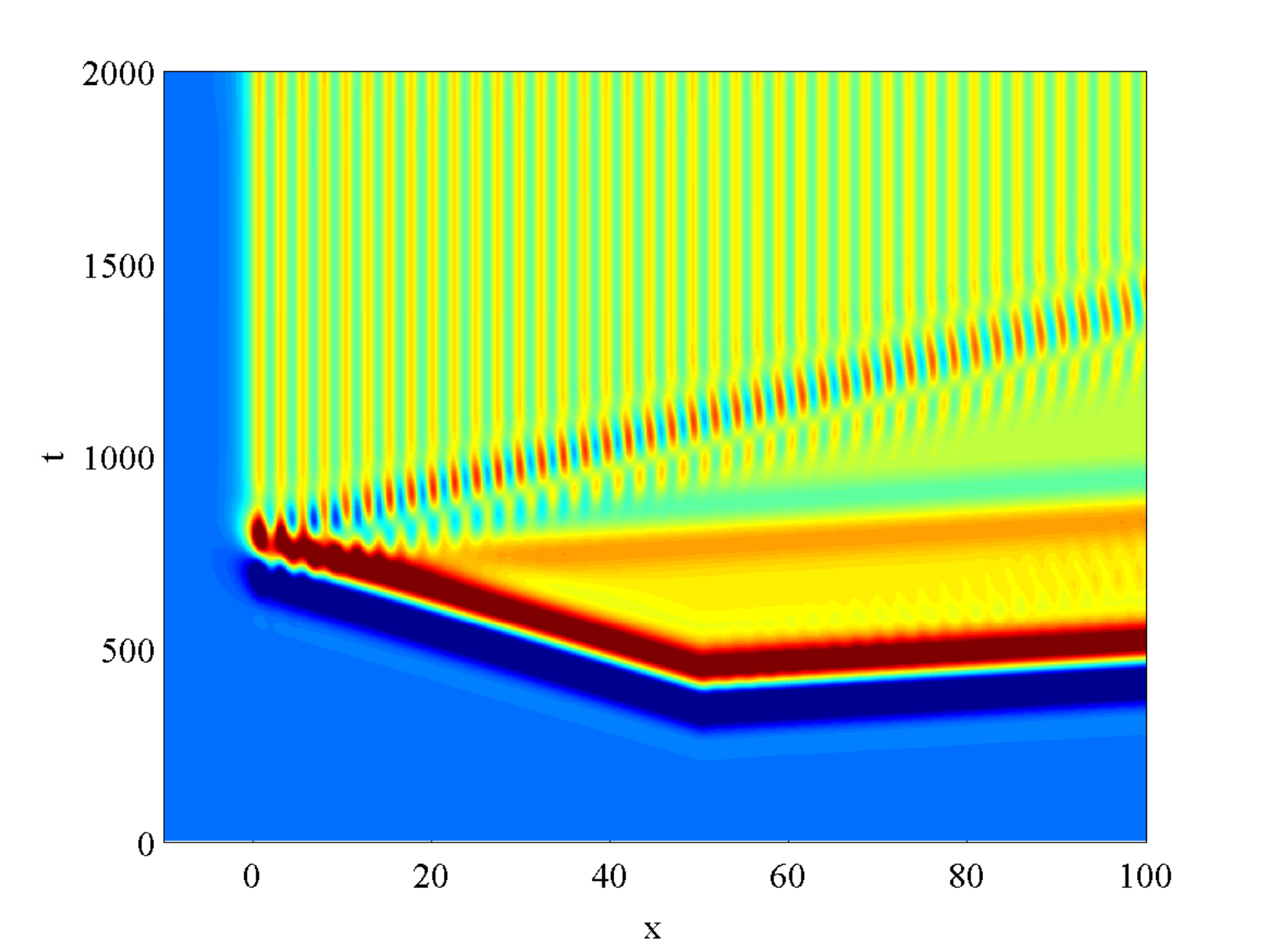}
\includegraphics[width=\linewidth]{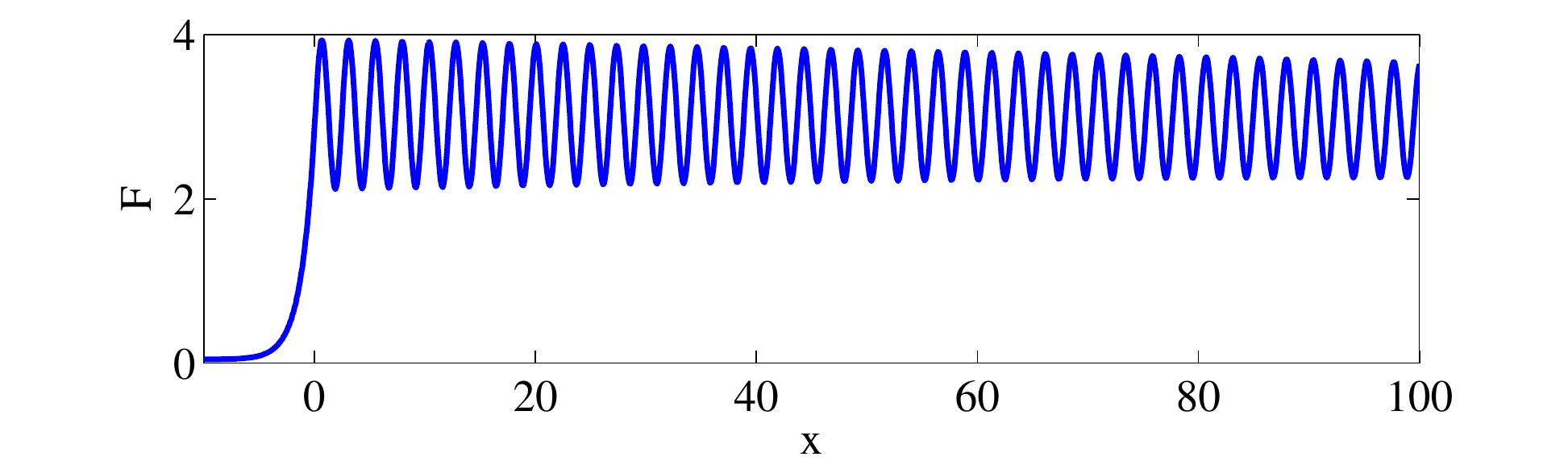}
\caption{Space-time 
plot of the undulation obtained when sending a wave packet initially centered around $x = 50$
in a white hole flow. The sonic horizon is located around $x= 0$. 
As in the main text, the sign of $M$ is positive and the subsonic region is on the right side of the horizon. 
The amplitude represented is that of $F = \int_0^1 \phi dy = \Phi + \frac{1}{3} \Psi$. The top plot shows the time-evolution with a linear color scale, and the bottom one shows the late-time configuration.} \label{fig:undul}
\end{figure}

\noindent
In \fig{fig:undul} we show a space-time diagram of the perturbation obtained by sending a localized wave-packet on a white hole horizon. The important point is the appearance of a long-lasting {\it undulation}, i.e., of a zero-frequency mode with a macroscopic amplitude. Its presence is due to the diverging character of $\left\lvert \beta_\om \right\rvert^2 \sim \left\lvert \alpha_\om \right\rvert^2$ as $1/ \om$ for $\om \to 0$ which amplifies the small-frequency components of the incident wave-packet. See Eq.~(23) in~\cite{Coutant:2012mf} and Eq.~(27) in~\cite{Busch:2014hla} for studies of the same mechanism in related contexts. 

\end{document}